\documentstyle{article}[12pt]
\newcommand{\nin}{\noindent}
\newcommand{\be}{\begin{equation}}
\newcommand{\ee}{\end{equation}}
\newcommand{\bea}{\begin{eqnarray}}
\newcommand{\eea}{\end{eqnarray}}
\newcommand{\br}{\hskip .25cm/\hskip -.25cm}
\newcommand{\hf}{\frac{1}{2}}
\newcommand{\nonu}{\nonumber\\}

\newcommand{\ol}{\overline}

\begin{document}

\begin{center}

{\Large{\bf An alternative approach to dynamical mass generation in $QED_3$}}

\vspace{1cm}

{\bf Jean Alexandre}\\
Physics Department, King's College \\
WC2R 2LS, London, UK\\
jean.alexandre@kcl.ac.uk

\vspace{3cm}

{\bf Abstract}

\end{center}

\vspace{.2cm}

\nin Some quantum properties of $QED_3$ are studied with the help of an exact evolution
equation of the effective action with the bare fermion mass. The resulting effective 
theory and the occurrence of a dynamical mass are discussed in the framework of the 
gradient expansion.

\vspace{3cm}

\section{Introduction}

$QED$ in 2+1 dimensions ($QED_3$) was introduced as a toy model for
chiral symmetry breaking \cite{pisarski,appelquist}. 
There the Schwinger-Dyson equation for the fermion propagator was solved,
in order to study the dynamical fermion mass generation. 
More recently, similar studies were done analytically and numerically \cite{maris},
where different resummations are
considered in the study of the Schwinger-Dyson equation, taking into account different
ans\"atze for the the vertex. 

Besides the toy model motivation, $QED_3$ happens to be relevant for 
the effective description of planar antiferromagnetic systems \cite{condmat}.  
Indeed, the $d-$wave pairing that occurs in these systems allows a linearization 
of the dispersion relation around the nodes in the energy gap, which leads to relativistic-like
effective theories.

Motivated by these recent developments, an alternative approach to the quantum
properties of $QED_3$ is proposed here. 
This approach is based on an exact functional method, where the quantum fluctuations are 
controlled via the bare fermion mass. When this mass is large, the interaction can 
be neglected and the quantum corrections are weak: the system is almost classical. 
As the bare mass decreases, the quantum
fluctuations gradually appear and the system sees the generation of quantum effects, leading to the 
the full quantum effective theory.

This functional method was also used in 3+1 dimensions for a scalar
theory \cite{intscal} and for $QED$ \cite{intqed},
in the description of a coulomb gas \cite{intmat} and in a Wess-Zumino model in 2+1
dimensions \cite{intsusy}. In all these cases, the usual renormalized parameters were
recovered at one-loop. Beyond one-loop, the resummation provided by the 
gradient expansion used in this functional method does not coincide anymore with the loop expansion.
In the example of $QED$ and for the one-loop approximation, a singularity appears in 
the coupling constant flow as the bare fermion mass decreases, i.e. as the fluctuations 
appear. This singularity is the analogue of the famous Landau pole and it disappears
beyond the one-loop approximation, i.e. when taking into account the full equations
obtained in this method.

In section 2, we set up the frame of this method and give the resulting flows that are
obtained, the computations being given in Appendix A and Appendix B. We show then in section 3
that the resulting effective couplings are consistent, at one-loop, with the usual 
renormalized parameters computed with standard Feynman graphs. In section 4, we give
the non-perturbative solution and discuss the generation of a dynamical mass in the 
massless case ("massless" refers here to the bare theory). 
A self-consistent relation is found between the effective coupling and the dynamically
generated mass. This relation shows that the only way to have a vanishing dynamical mass
is to have a non-interacting effective theory.
This result is set in the lowest order of the gradient expansion and
thus the influence of the number of flavours is not discussed, as explained in section 5
which is let for a discussion.

\section{Evolution equations}

We consider a parity conserving theory, what is possible with $2N$ fermions,
constituing $N$ couples in which the masses are opposite.
The $2N$ fermions are then written as $N$ 4-component 
fermions and we consider the $4\times 4$ reducible representation of the $\gamma$-matrices 
given in \cite{appelquist}. 

The starting point is the bare action 

\bea
S&=&\int d^3x \Big\{\hf A^\mu(T_{\mu\nu}+\alpha L_{\mu\nu})A^\nu\\
&&~~~~+\sum_{k=1}^N \ol\psi_k(i\br\partial-g_B\br A-\lambda m_B)\psi_k\Big\}\nonumber,
\eea

\nin where $T_{\mu\nu}$ and $L_{\mu\nu}$ are respectively the transverse and longitudinal parts
of the photon propagator and $\alpha$ is a usual gauge-fixing parameter.

The control of the quantum effects consists in playing with $\lambda$. As 
this parameter goes to $\infty$, the fermion bare mass is large and thus freezes the 
quantum fluctuations: the system is classical and the effective theory is the bare one.
On the other hand, as $\lambda$ decreases, the quantum corrections gradually appear in the
effective theory. 

The idea of the method is to compute the evolution of the effective action $\Gamma$ 
(the proper graphs generator functional) with $\lambda$.
Starting from $\lambda=\infty$, where $\Gamma=S$, the massive case
will be obtained when $\lambda$ decreases down to 1 (see section 3) and the massless
case when $\lambda$ decreases down to 0 (see section 4).

The evolution equation for $\Gamma$ was derived for $N=1$ in \cite{intqed} and
we review in the Appendix A this derivation for any $N$. The result reads

\be\label{evolequa}
\partial_\lambda\Gamma=m_B\mbox{Tr}\sum_{k=1}^N\left\{-\ol\psi_k\psi_k
+i\left(\frac{\delta^2\Gamma}{\delta\ol\psi_k\delta\psi_k}\right)^{-1}\right\},
\ee

\nin and we recall that this evolution equation is exact and does not rely on any approximation.
In the framework of the gradient expansion, we then make the projection of $\Gamma$ on the 
following functional subspace:

\bea\label{gradexp}
\Gamma&=&\int d^3x\Big\{\hf A^\mu(\beta_T T_{\mu\nu}+\alpha\beta_L L_{\mu\nu})A^\nu\\
&&~~~~+\sum_{k=1}^N \ol\psi_k\left(iz\br\partial-g z\br A-m\right)\psi_k\Big\}\nonumber,
\eea

\nin where $\beta_T,\beta_L,z,g,m$ are functions of $\lambda$. 
The step (\ref{gradexp}) constitutes our approximation and is valid in the IR. 
It is shown in the Appendix B that plugging this ansatz in the evolution 
equation (\ref{evolequa}) leads to the following differential equations:

\bea\label{equainit}
\partial_\lambda\beta_L&=&0\\
\partial_\lambda m&=&m_B\nonu
\partial_\lambda\beta_T&=&-N\frac{g^2 z m_B}{6\pi m^2}\nonu
\partial_\lambda z&=&\frac{g^2 z^2 m_B}{6\pi m^2}\frac{1}{\alpha\beta_L}\nonu
\partial_\lambda (gz)&=&\frac{g^3 z^2 m_B}{18\pi m^2}\left(\frac{2}{\beta_T}+
\frac{1}{\alpha\beta_L}\right)\nonumber
\eea

\nin It is interesting to note the following points:

\begin{itemize}

\item As expected, the photon does not acquire a mass in the evolution with $\lambda$.

\item The longitudinal part $\beta_L$ of the photon propagator does not evolve with $\lambda$.

\item In the Landau gauge $(\alpha=\infty)$, there is no fermion wave function
renormalization and $z=1$.

\item The fermion mass $m$ has a classical flow with $\lambda$ (no quantum contribution).

\end{itemize}

\nin As a consequence, we have 

\bea
\beta_L&=&1,\\
m&=&m_B(\lambda+c)\nonumber,
\eea

\nin where $c$ is a constant of integration linked to the dynamical mass generated in the
massless case since.

\be
m_{dyn}=\lim_{\lambda\to 0} \left(\frac{m}{z}\right)=c\frac{m_B}{z_0},
\ee

\nin where $z_0=z(0)$. 

If we note $\beta_T=\beta$, the remaining differential equations can be written

\bea
\partial_\lambda\beta&=&-N\frac{g^2 z}{6\pi m_B}\frac{1}{(\lambda+c)^2}\\
\partial_\lambda z&=&\frac{1}{\alpha}\frac{g^2 z^2}{6\pi m_B}\frac{1}{(\lambda+c)^2}\nonu
\partial_\lambda (gz)&=&\frac{g^3 z^2}{18\pi m_B}\frac{1}{(\lambda+c)^2}
\left(\frac{2}{\beta}+\frac{1}{\alpha}\right)\nonumber
\eea

\nin Finally, we will consider the Landau gauge where $z=1$ and the differential equations 
that remain read then

\bea\label{equadifflandau}
\partial_\lambda\beta&=&-N\frac{g^2}{6\pi m_B}\frac{1}{(\lambda+c)^2}\\
\partial_\lambda g&=&\frac{g^3}{9\pi\beta m_B}\frac{1}{(\lambda+c)^2}\nonumber.
\eea

\nin In this case, the dynamical mass is given by

\be\label{mdyn} 
m_{dyn}=cm_B.
\ee

\nin We remind that the bare fermion mass is $\lambda m_B$ and not $m_B$, such that the 
massless case is reached when $\lambda\to 0$. $m_B$ must be seen as an initial parameter
which cannot vanish (the whole method is based on $m_B\ne 0$!). The 
possibility of having $m_{dyn}=0$ then occurs if $c=0$.

\section{One-loop solution (massive case)}

To obtain the renormalized parameters in the massive case, we integrate the
differential equations (\ref{equadifflandau}) from $\lambda=\infty$ where the quantum corrections are frozen
$(g(\infty)=g_B,\beta(\infty)=1)$, to $\lambda=1$ where all the quantum effects are generated
(we note $g(1)=g_1,\beta(1)=\beta_1$).

Before going to the full treatment of the differential equations (\ref{equadifflandau}), let us check
that these lead to the expected results at one-loop. We note for this that the right-hand sides
of Eqs.(\ref{equadifflandau})
are proportional to $\hbar$, such that the one-loop approximation consists in replacing the
parameters of the right-hand sides by the bare ones (and thus take $c\to 0$ since this constant arises 
from quantum effects). We have then, restoring the factors $\hbar$,

\bea
\partial_\lambda\beta&=&-\hbar N\frac{g_B^2}{6\pi m_B}\frac{1}{\lambda^2}+...\\
\partial_\lambda g&=&\hbar\frac{g_B^3}{9\pi m_B}\frac{1}{\lambda^2}+...\nonumber,
\eea

\nin where the dots stand for higher orders in $\hbar$. The integration of these equations from 
$\lambda=\infty$ to $\lambda=1$ gives then

\bea\label{oneloop}
\beta_1&=&1+\hbar N\frac{g_B^2}{6\pi m_B}+...\\
g_1&=&g_B-\hbar\frac{g_B^3}{9\pi m_B}+...\nonumber.
\eea

\nin It can be checked with standard computations that these results are given by the one-loop
Feynman graphs. Indeed, the graph corresponding to the one-loop vacuum polarization 
(for one fermion flavour) is

\bea
&&i(-ig_B)^2tr\int_p\gamma^\mu\frac{\br p+m_B}{p^2-m_B^2}\gamma^\nu
\frac{(\br p-\br k)+m_B}{(p-k)^2-m_B^2}\\ 
&=&-i\frac{g^2_Bg^{\mu\nu}}{2\pi^2}\int_0^1 dx\int d^3q\frac{m_B^2-q^2-x(1-x)k^2+2q^2/3}
{[q^2-m_B^2+x(1-x)k^2]^2}\nonu
&&+\frac{4g_B^2}{\pi^2}k^2T^{\mu\nu}(k)\int_0^1 dx\int_0^\infty dq_E
\frac{x(1-x)q_E^2}{[q_E^2+m_B^2-x(1-x)k^2]^2}\nonu
&=&\frac{g_B^2}{6\pi m_B}k^2T^{\mu\nu}(k)+\mbox{higher orders in }k\nonumber,
\eea

\nin where $q=p-(1-x)k$ and $q_E$ denotes the Euclidean momentum. As expected, 
the would-be linearly divergent integral (second line) cancels
when using a gauge invariant regularization. 
The one-loop function $\beta_1$ for $N$ flavours is then given by Eq.(\ref{oneloop}). 
Note that this result is consistent with the limit $k<<m_B$ of the large-$N$ polarization
tensor \cite{pisarski}.

The graph corresponding to the one-loop correction
to the coupling is, in the Landau gauge and in the limit of zero incoming momenta,

\bea
&&(-ig_B)^3\int_p\frac{1}{p^2}T_{\mu\nu}(p)\gamma^\mu\frac{\br p+m_B}{p^2-m_B^2}\gamma^\rho
\frac{\br p+m_B}{p^2-m_B^2}\gamma^\nu\\
&=&-\frac{2g_B^3}{9}\gamma^\rho\int_{p_E}\frac{1}{p^2_E}\frac{p_E^2+3m_B^2}{(p_E^2+m_B^2)^2}\nonu
&=&-\frac{g_B^3}{9\pi m_B}\gamma^\rho\nonumber,
\eea

\nin which leads to the result in Eq.(\ref{oneloop}).

\section{Non-perturbative solution}

We turn now to the complete resolution of the differential equations (\ref{equadifflandau}).
Taking the ratio of these two equations and integrating, we have
for any $\lambda$

\be\label{betag}
\beta=\left(\frac{g_B}{g}\right)^{3N/2}.
\ee

\nin Plugging this result in the differential equation (\ref{equadifflandau}) for $g$ gives

\be
g^{-3-3N/2}\partial_\lambda g=\frac{g_B^{-3N/2}}{9\pi m_B}\frac{1}{(\lambda+c)^2}.
\ee

\nin The integration from $\lambda=\infty$ down to any $\lambda$ leads to

\be\label{g1gB}
g(\lambda)=g_B\left(1+\frac{3N+4}{18\pi}\frac{g_B^2}{m(\lambda)}\right)^{-2/(3N+4)}.
\ee 

\nin The solution for the photon propagator renormalization is then 

\be
\beta(\lambda)=\left(\frac{g_B}{g(\lambda)}\right)^{3N/2}=
\left(1+\frac{3N+4}{18\pi}\frac{g_B^2}{m(\lambda)}\right)^{3N/(3N+4)}.
\ee

So as to discuss the dynamical mass generation,  
we should take the limit $\lambda\to 0$, where
$m(0)=m_{dyn}$ and we note $g(0)=g_0$. We have then

\be\label{equafinal}
g_0=g_B\left(1+\frac{3N+4}{18\pi}\frac{g_B^2}{m_{dyn}}\right)^{-2/(3N+4)}
\ee

\nin From this last equation we see that the only possibility to have $m_{dyn}=0$ is $g_0=0$.
It is indeed expected that no dynamical mass can be generated in a non-interacting effective theory.
Eq.(\ref{equafinal}) should be seen as a self-consistent 
relation between the renormalized parameters $g_0$ and $m_{dyn}$, relation which is parametrized by $g_B$.
This result, valid in the present approximation of the
gradient expansion, is independent of $N$.

\section{Discussion}

Let us first show the link between the equations obtained by this method and the
Schwinger-Dyson equation for the fermion propagator.

One can see from the Appendix B that the evolution equation of the fermion propagator
is 

\bea
\partial_\lambda G^{-1}(k)&=&-m_B-im_Bg^2z^2\int_p\gamma_\mu G^2(p)
\gamma_\nu D^{\mu\nu}(k+p)\nonu
&=&-m_B-im_Bg_B^2\int_p\gamma_\mu G_B^2(p)\gamma_\nu D_B^{\mu\nu}(k+p)\nonu
&&~~~~~~~~~~~~~~+\mbox{higher orders in }\hbar, 
\eea

\nin where $G_B$ and $D_B^{\mu\nu}$ are the bare fermion 
and photon propagators respectively. But since $G_B^{-1}(k)=\br k-\lambda m_B$, one can note that

\be
\partial_\lambda G_B(k)=m_BG_B^2(k),
\ee

\nin such that the one-loop approximation to the evolution equation of $G^{-1}$ 
can be written

\be
\partial_\lambda G^{-1}(k)=\partial_\lambda G^{-1}_B(k)-ig_B^2\int_p
\gamma_\mu\partial_\lambda G_B(p)\gamma_\nu D_B^{\mu\nu}(k+p).
\ee

\nin Since $D_B^{\mu\nu}$ is independent of $\lambda$, the previous equation integrates as

\be
G^{-1}(k)=G_B^{-1}(k)-ig_B^2\int_p\gamma_\mu G_B(p)\gamma_\nu D_B^{\mu\nu}(k+p).
\ee

\nin Therefore the one-loop approximation to the evolution equation of the fermion propagator
gives the usual one-loop self energy, which is also the one-loop truncation of 
the Schwinger-Dyson equation for the fermion propagator. 
The higher-order terms in $\hbar$ are different though, since the resummations provided by 
the two methods are not the same. 

The study of dynamical mass generation with the Schwinger-Dyson method \cite{pisarski,appelquist,maris}
takes into account the momentum dependence of the
fermion and photon self energies, what is not the case with the gradient expansion truncated to the
present IR approximation. For this reason, the influence of the number of flavours is not
discussed here. Indeed, it is the momentum-dependence 
of the fermion self energy which leads to a condition on the number of flavours: it was found in 
\cite{appelquist} that there is no dynamical mass generated above $N_c=32/\pi^2$, and the studies
made in \cite{maris} agree with the existence of $N_c$, and of the same order.

The next step thus, to discuss the influence of the number of flavours,  would be to go further in 
the gradient expansion, so as to take into account
the momentum dependence of the self energies. But this would lead to "partial-integro-differential"
equations (derivative with respect to $\lambda$ and integration with respect to the momentum)
and would involve new approximations, similar to the ones used in the study of 
Schwinger-Dyson equations. The present scheme is based on the gradient expansion approximation {\it only} and
is therefore believed to give a relevant insight into this problem of dynamical mass generation, besides 
the study of the role of $N$.

To obtain an explicit expression of the dynamical mass as a function of the bare coupling only
(and therefore compute the value of the constant $c$ in Eq.(\ref{mdyn})),
one would need to take a finite initial value for $\lambda$, corresponding to a system where 
some quantum fluctuations are already generated. The initial values of the parameters defining 
the theory could then be taken from a loop-expansion, valid when $\lambda>>1$. But the result would then 
depend on the initial value of $\lambda$, whereas the motivation of the present work 
was to find a self-consistent equation like (\ref{equafinal}), independent 
of the $\lambda$-scheme used here. 

\vspace{1cm}

\nin{\bf Acknowledgements:} I would like to thank Janos Polonyi for useful discussions.

\section*{Appendix A: Evolution equation for $\Gamma$}

The bare action, functional of the fields $\hat{\ol\psi},\hat\psi,\hat A_\mu$ is

\bea
S&=&\int d^3x \Big\{\hf \hat A^\mu(T_{\mu\nu}+\alpha L_{\mu\nu})\hat A^\nu\\
&&~~~~+\sum_{k=1}^N \hat{\ol\psi}_k(i\br\partial-g_B\br{\hat A}-\lambda m_B)\hat\psi_k\Big\}\nonumber,
\eea

\nin and the connected graphs generator functional, function of the sources $\eta_k,\ol\eta_l,j_\mu$
is defined by $W=-i\ln{\cal Z}$ where

\be
{\cal Z}=\int{\cal D}[\hat{\ol\psi},\hat\psi,\hat A_\mu]
\exp\left\{iS+i\int d^3x\left(\j^\mu \hat A_\mu+\sum_{k=1}^N(\hat{\ol\psi}_k\eta_k+
\ol\eta_k\hat\psi_k)\right)\right\}.
\ee

\nin The functional derivatives of $W$ define the expectation value fields $\ol\psi_k,\psi_l,A_\mu$:

\bea\label{defquant}
\frac{\delta W}{\delta\ol\eta_l}&=&\frac{1}{{\cal Z}}<\hat\psi_l>=\psi_l\nonu
\frac{\delta W}{\delta\eta_k}&=&-\frac{1}{{\cal Z}}<\hat{\ol\psi}_k>=-\ol\psi_k\nonu
\frac{\delta W}{\delta j_\mu}&=&\frac{1}{{\cal Z}}<\hat A_\mu>=A_\mu,
\eea

\nin where for any $\hat\phi$

\be
<\hat\phi>=\int{\cal D}[\hat\phi]~\hat\phi~
\exp\left\{iS+i\int d^3x\left(j^\mu\hat A_\mu+\sum_{k=1}^N(\hat{\ol\psi}_k\eta_k+\ol\eta_k\hat\psi_k)\right)\right\}.
\ee

\nin We also have

\be\label{d2W}
\frac{\delta^2 W}{\delta\ol\eta_k\delta\eta_l}
=i\ol\psi_l\psi_k-\frac{i}{Z}\left<\hat{\ol\psi}_l\hat\psi_k\right>.
\ee

\nin Inverting the relations (\ref{defquant}) which give the expectation value fields 
as functions of the sources,
we define the Legendre transform $\Gamma$ of $W$ by

\be
\Gamma=W-\int d^3x\left\{j^\mu A_\mu+\sum_{k=1}^N(\ol\psi_k\eta_k+\ol\eta_k\psi_k)\right\}.
\ee

\nin From this definition we extract the following functional derivatives:

\bea\label{d2G}
\frac{\delta\Gamma}{\delta\psi_l}&=&\ol\eta_l\\
\frac{\delta\Gamma}{\delta\ol\psi_k}&=&-\eta_k\\
\frac{\delta\Gamma}{\delta A^\mu}&=&-j_\mu\\
\frac{\delta^2\Gamma}{\delta\ol\psi_k\delta\psi_l}&=&
-\left(\frac{\delta^2 W}{\delta\ol\eta_l\delta\eta_k}\right)^{-1}\nonumber
\eea

\nin The evolution of $W$ with the parameter $\lambda$ is, according to (\ref{d2W}),

\bea
\partial_\lambda W&=&-\frac{m_B}{{\cal Z}}
\int d^3x\sum_{k=1}^N <\hat{\ol\psi}_k\hat\psi_k>\nonu
&=&-m_B\int d^3x\sum_{k=1}^N\ol\psi_k\psi_k
-im_B\int d^3x\sum_{k=1}^N\frac{\delta^2 W}{\delta\ol\eta_k\eta_k}.
\eea

\nin To compute the evolution of $\Gamma$ with $\lambda$, we have to keep in mind that its
independent variables are $\ol\psi_k,\psi_l,A_\mu$ and $\lambda$, such that

\bea
\partial_\lambda \Gamma&=&\partial_\lambda W+
\int d^3x\left\{ \frac{\delta W}{\delta j_\mu}\partial_\lambda j_\mu
+\sum_{k=1}^N\left(-\frac{\delta W}{\delta\eta_k}\partial_\lambda\eta_k 
+\partial_\lambda\ol\eta_k\frac{\delta W}{\delta\ol\eta_k}\right)\right\}\nonu
&&-\int d^3x \left\{\partial_\lambda j^\mu A_\mu
+\sum_{k=1}^N\left(\ol\psi_k\partial_\lambda\eta_k+\partial_\lambda\ol\eta_k\psi_k\right)\right\}\nonu
&=&\partial_\lambda W.
\eea

\nin Combining these different results, we finally obtain the exact evolution equation for
the proper graphs generator functional $\Gamma$:

\bea\label{evol}
&&\partial_\lambda\Gamma=
m_B\int d^3x\sum_{k=1}^N\left\{-\ol\psi_k\psi_k+
i\left(\frac{\delta^2\Gamma}{\delta\ol\psi_k\delta\psi_k}\right)^{-1}\right\}\\
&=&m_B\int_{pq}\delta^3(p+q)
\sum_{k=1}^N\left\{-\ol\psi_k(p)\psi_k(q)
+i\left(\frac{\delta^2\Gamma}{\delta\ol\psi_k(p)\delta\psi_k(q)}\right)^{-1}\right\},\nonumber
\eea

\nin where we denote

\be
\int_p(...)=\int\frac{d^3p}{(2\pi)^3}(...).
\ee

\section*{Appendix B: Evolution equations for the parameters}

We decompose the second derivative of $\Gamma$ in a sum of the field-independent diagonal part 
$\Delta$ and the part $\Theta$ which contains the fields. The inverse of $\delta^2\Gamma$ is then taken as 
the following expansion

\bea
(\delta^2\Gamma)^{-1}&=&\Delta^{-1}-\Delta^{-1}\Theta\Delta^{-1}+\Delta^{-1}\Theta\Delta^{-1}\Theta\Delta^{-1}\nonu
&&~~~~-\Delta^{-1}\Theta\Delta^{-1}\Theta\Delta^{-1}\Theta\Delta^{-1}+...,
\eea 

\nin and this expansion is truncated up to the desired order consistent with the approximation (\ref{gradexp})
that we consider, i.e. up to the order 3 in the non-diagonal part $\Theta$. The result is then (the flavor indices 
are not written)

\bea\label{d2Gamma}
\mbox{Tr}\left(\frac{\delta^2\Gamma}{\delta\ol\psi\delta\psi}\right)^{-1}&=&
\mbox{Tr}\Big\{G+G\Big(-\phi+\ol\theta^\mu D_{\mu\nu}\theta^\nu+\phi G\phi\nonu
&&~~~~-2\ol\theta^\mu D_{\mu\nu}\theta^\nu G\phi-\phi G\phi G\phi+...\Big)G\Big\},
\eea

\nin where (in Fourier components)

\bea
G(p,q)&=&\frac{z\br p+m}{z^2p^2-m^2}\delta^3(p+q)\nonu
D_{\mu\nu}(p,q)&=&\frac{1}{p^2}\left(\frac{1}{\beta_T}T_{\mu\nu}(p)+
\frac{1}{\alpha\beta_L}L_{\mu\nu}(p)\right)\delta^3(p+q)\nonu
\phi(p,q)&=&-zg\br A(-p-q)\nonu
\theta^\mu(p,q)&=&-zg\gamma^\mu\psi(-p-q)\nonu
\ol\theta^\nu(p,q)&=&zg\ol\psi(-p-q)\gamma^\nu,
\eea

\nin with 

\bea
T_{\mu\nu}(p)&=&g_{\mu\nu}-\frac{p_\mu p_\nu}{p^2}\nonu
L_{\mu\nu}(p)&=&\frac{p_\mu p_\nu}{p^2}.
\eea

\nin The contribution of Eq.(\ref{d2Gamma}) to the fermion propagator is then

\bea
&&\mbox{Tr}\left\{G\ol\theta^\mu D_{\mu\nu}\theta^\nu G\right\}\\
&&=-g^2z^2\int_k\ol\psi(k)
\left(\int_p\gamma^\nu G^2(p)\gamma^\mu D_{\mu\nu}(k+p)\right)\psi(-k)\nonu 
&&=-i\frac{g^2z^2}{6\pi m^2}\frac{1}{\alpha\beta_L}\int_k\ol\psi(k)\br k\psi(-k)
+\mbox{higher orders in }k\nonumber,
\eea

\nin where the higher orders in $k$ are not considered in the present approximation
(\ref{gradexp}) of the gradient expansion.
We note that the mass term (order $k^0$) vanishes.
The contribution of Eq.(\ref{d2Gamma}) to the photon propagator is

\bea
&&\mbox{Tr}\left\{G\phi G\phi G\right\}\\
&&=-g^2z^2\int_k A_\mu(-k)
\left(\mbox{tr}\int_p G^2(p)\gamma^\mu G(k+p)\gamma^\nu\right)A_\nu(k)\nonu
&&=i\frac{g^2z^2}{12\pi m^2 z}\int_k A_\mu(-k)A_\nu(k)k^2 T^{\mu\nu}(k)
+\mbox{higher orders in }k\nonumber.
\eea

\nin As expected, no photon mass is generated and 
the correction to the photon propagator is transverse.
Finally, the contribution of Eq.(\ref{d2Gamma}) to the vertex is

\bea
&&\mbox{Tr}\left\{-2G\ol\theta^\mu D_{\mu\nu}\theta^\nu G\phi G\right\}\\
&&=-2g^3 z^3\int_{kp}
\ol\psi(k)A_\rho(-p)\left(\int_q\gamma^\mu G(q)\gamma^\rho G^2(p+q)\gamma^\nu 
D_{\mu\nu}(k+q)\right)\psi(p-k)\nonu
&&=i\frac{g^3 z^2}{18\pi m^2}\left(\frac{2}{\beta_T}+\frac{1}{\alpha\beta_L}\right)
\int_{kp}\ol\psi(k)\br A(-p)\psi(p-k)\nonu
&&~~~~+\mbox{higher orders in {\it k} and {\it p}}\nonumber.
\eea

\nin The last contribution $G\phi G\phi G\phi G$ in Eq.(\ref{d2Gamma}) gives a cubic term in the photon 
field and vanishes by symmetry.
The identification of the right-hand side and left hand side of the evolution equation
(\ref{evolequa}) leads then to the differential equations (\ref{equainit}).

\end{document}